\documentclass[twocolumn,superscriptaddress,amsmath, amssymb, aps, prl]{revtex4-1}
\usepackage{color,soul}
\usepackage{graphicx}
\usepackage{bm}
\usepackage{dcolumn}
\usepackage{sidecap}
\usepackage{amsmath}
\usepackage[version=4]{mhchem}
\usepackage{etoolbox}
\apptocmd{\sloppy}{\hbadness 10000\relax}{}{}

\newcommand{\nH}{n_{\textsc{\tiny \emph{H}}}}
\newcommand{\SZ}{\langle S_Z \rangle}
\newcommand{\LZ}{\langle L_Z \rangle}
\newcommand{\TZ}{\langle T_Z \rangle}
\newcommand{\JZ}{\langle J_Z \rangle}

\newcommand*{\citen}{}
\DeclareRobustCommand*{\citen}[1]{%
  \begingroup
    \romannumeral-`\x 
    \setcitestyle{numbers}%
    \cite{#1}%
  \endgroup
}

\begin{document}

\title{Intercalated Europium Metal in Epitaxial Graphene on SiC}

\author{Nathaniel A. Anderson}
\affiliation{Ames Laboratory and Department of Physics and Astronomy, Iowa State University, Ames, Iowa 50011, USA}

\author{Myron Hupalo} 
\affiliation{Ames Laboratory and Department of Physics and Astronomy, Iowa State University, Ames, Iowa 50011, USA}

\author{David Keavney}
\affiliation{X-ray Science Division, Advanced Photon Source, Argonne National Laboratory, Lemont, Illinois 60439, USA}

\author{Michael C. Tringides}
\affiliation{Ames Laboratory and Department of Physics and Astronomy, Iowa State University, Ames, Iowa 50011, USA}

\author{David Vaknin}
\email{vaknin@ameslab.gov}
\affiliation{Ames Laboratory and Department of Physics and Astronomy, Iowa State University, Ames, Iowa 50011, USA}

\date{\today}

\begin{abstract}
X-ray magnetic circular dichroism (XMCD) reveal the magnetic properties of intercalated europium metal under graphene on SiC(0001).  Intercalation of Eu nano-clusters (average size 2.5 nm) between graphene and SiC substate are formed by deposition of  Eu on epitaxially grown graphene that is subsequently annealed at various temperatures while keeping the integrity of the graphene layer. Using sum-rules analysis of the XMCD of Eu M$_{4,5}$ edges at  $T = 15$ K, our samples show paramagnetic-like behavior with distinct anomaly at T $\approx$ 90 K which may be related to the N{\`e}el transition, T$_N$ = 91 K, of bulk metal Eu. We find no evidence of ferromagnetism due to EuO or antiferromagnetism due to \ce{Eu2O3} indicating that the graphene layer protects the intercalated metallic Eu against oxidation over months of exposure to atmospheric environment.
\end{abstract}
\maketitle

\section*{Introduction}
In addition to its unique electronic properties and optical transparency, that render it potential applications in spintronics and photovoltaic devices\cite{Zhang2014a}, graphene has been recognized as the ultimate mono-atomic protective membrane of metal surfaces against corrosion\cite{Dedkov2008a,Dedkov2008,Bohm2014,Bunch2008}. The chemical vapor deposition (CVD) of graphene has by now been established as a scalable method for depositing graphene albeit with inevitable surface defects due to the non-epitaxial nature of the growth that proceeds at multiple points of nucleation\cite{Hofmann2015}. So, covering weak oxidizing metals (e.g. Ni, Co) with graphene can protect their surfaces over long periods to atmospheric exposure\cite{Weatherup2015}, because the metal-oxides formed at defects protect against further oxidation. On the other hand, for strong oxidizers (e.g., Fe or Eu) in atmospheric environment, corrosion through graphene-defects or other protective layers gradually spreads over the whole surface and even penetrates  the bulk\cite{Weatherup2015,Anderson2017a}.  Defect-free and epitaxial monoatomic layer of graphene has long been produced on SiC(0001) forming a continuous membrane over the whole surface including surface steps\cite{Forbeaux1998}.  
With these advances, modifying the electronic properties of graphene has been investigated either by depositing inert metals\cite{ren2010,Forster2012} and metal-oxides\cite{Wang2008} or by intercalating between the graphene and the metal substrate\cite{Premlal2009,Schumacher2014,Voloshina2014,Huttmann2017} or between graphene and the SiC buffer-layer\cite{Sung2017}.  Intercalation of metal donors or molecular acceptors into graphite is an old topic that culminated in recipes that enable control of the superstructures (staging phenomena), electrical conduction, superconductivity and even electrical energy storage in batteries (i.e., CF and CLi$_6$)\cite{Dresselhaus1981}.  Thus, intercalation with magnetic metals is a route to modify interfacial magneto-electronic properties with  potential applications in spintronics. Here, we report on the magnetic properties of intercalated Eu atoms  between graphene and the SiC buffer-layer by employing synchrotron X-ray magnetic circular dichroism (XMCD).  We also, report on the chemical stability of the buried Eu layer as the sample is exposed to air over a period of months.  Recently, intercalation of Eu between Ir substrate and graphene (prepared by CVD) reveals that the structure and magnetic properties of the intercalated Eu depend on the coverage which does not seem to affect the electronic structure\cite{Schumacher2014}.  However, a recent study shows that Eu intercalation between graphene and the SiC buffer layer modifies the $\pi-$band of graphene significantly\cite{Sung2017}.  We note that besides the different substrates, the intercalated phases formed in SiC\cite{Sung2017} are of higher coverages than those reported on graphene/Ir metal\cite{Schumacher2014}.

\section*{Samples and Methods}
\begin{figure}
\centering \includegraphics [width = 76 mm]  {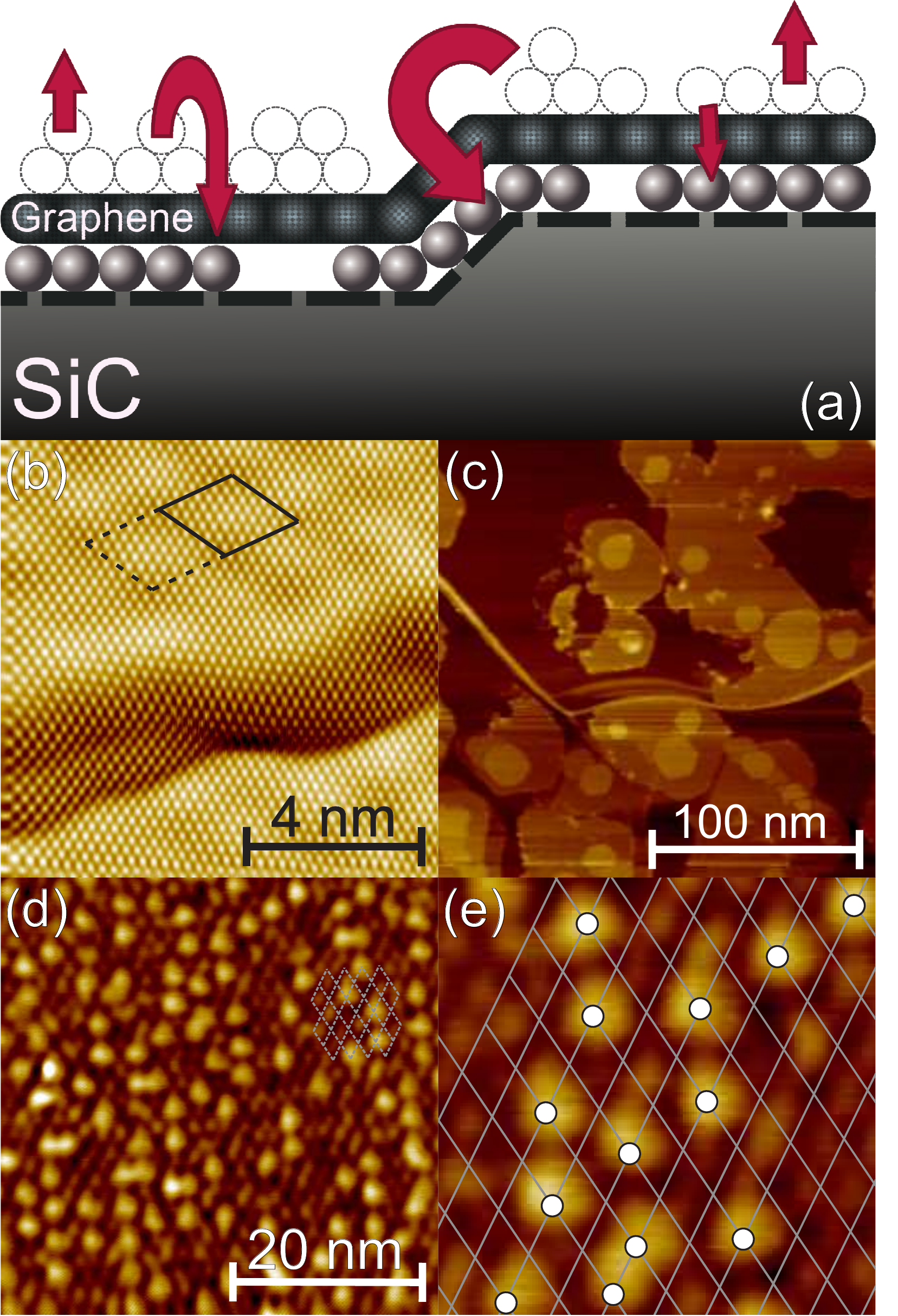}
\caption{(Color online) (a) Schematic illustration of intercalation after deposition of Eu metal on graphene.  During the annealing process, some atoms penetrate through the graphene and intercalate and some just evaporate.  (b) STM image of a pristine graphene on a SiC(0001) surface showing the well established diffuse 6$\times$6 superlattice. (c) deposited Eu islands on graphene before intercalation. (d) Eu intercalated under graphene forming 2-3 nm clusters.  (e) Eu clusters seem to randomly occupy the vertices on the superstructure grid.}
\label{fig:STM}
\end{figure}

\begin{figure}
\centering \includegraphics [width = 76 mm]  {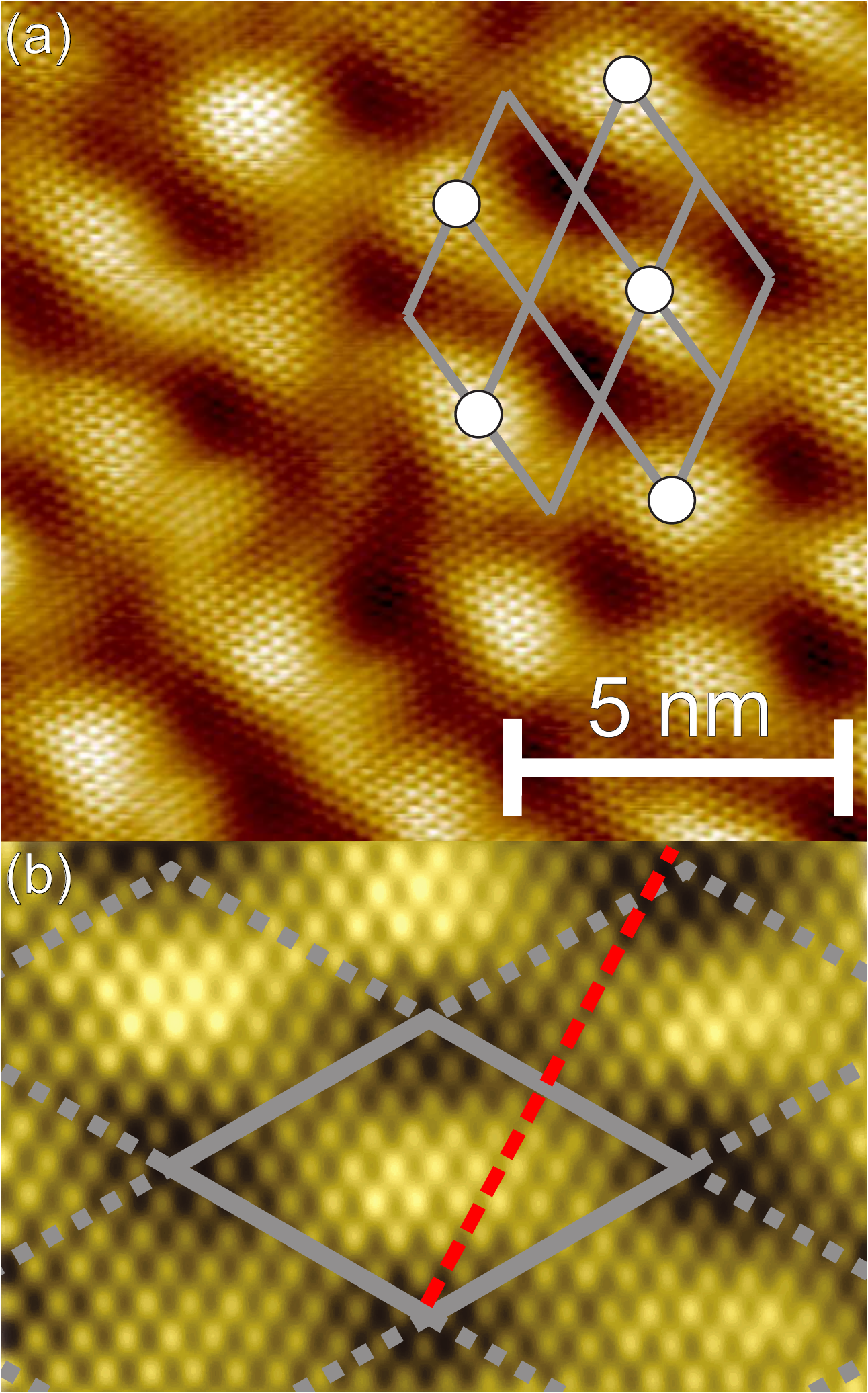}
\caption{(Color online) (a) STM image of a higher density cluster region showing that the graphene can still be seen on top of the clusters. The superstructure is rotated 30$^{\circ}$ with respect to the graphene lattice. (b) Enlarged and enhancedA region of the pristine graphene from Fig.\ \ref{fig:STM} showing that the superstructure from the buffer layer-SiC interface (solid diamond) is rotated 30$^{\circ}$ with respect to the graphene (dashed line). }
\label{fig:STMclose}
\end{figure}

The substrate used in our studies, 6H-SiC(0001) purchased from Cree, Inc.,  is graphitized in ultra-high vacuum (UHV, $P \approx1\cdot10^{-10}$ Torr) by direct current heating of the sample to $\sim$1300 C (measured with an infrared pyrometer). Figure\ \ref{fig:STM}b  shows a graphene layer with distinct 6$\times$6 superstructure commonly observed with graphene on SiC\cite{Forbeaux1998}. Metal intercalation is achieved by initial deposition of  nominal several monolayers of Eu metal on a SiC supported graphene (see Fig.\ \ref{fig:STM}c) followed by annealing, leading to two competing processes namely, intercalation/diffusion of metal atoms through the graphene sheet and atom desorption from the graphene surface into the vacuum (see illustration in Fig.\ \ref{fig:STM}a). Slow step-wise annealing up to the metal desorption temperature provides conditions preferred for intercalation.  After complete atom desorption, STM images show an undamaged graphene surface but with bright spots due to Eu clusters that are situated at the vertices of the 6$\times$6 superstructure (Fig.\ \ref{fig:STM}d-e, Fig.\ \ref{fig:STMclose}, and in the SI\cite{SI-EU}). The high-resolution STM images (Fig.\ \ref{fig:STMclose}) confirm that clusters are formed beneath the graphene, and that the cluster superstructure is rotated 30$^{\circ}$ with respect to the graphene. 
Under further prolonged annealing up to 1200 C, Eu atoms de-intercalate and the initial graphene interface can be restored. This indicates that the density of an intercalated metal can be controlled in intercalation/de-intercalation cycling.  
We note that lower annealing temperatures has been reported in the Ref.\ \cite{Sung2017} (120 C), confining the Eu diffusion between graphene and buffer layer, whereas annealing at 300 C shifts the Eu between the buffer and SiC and transforms the graphene to a bilayer. The annealing temperatures are higher in the current study resulting in a self-organized network of clusters of $\sim$25 atoms separated by at least 1.8 nm which behave independently in their magnetic response.

The location of the intercalated metal whether between graphene and buffer layer or between buffer layer and SiC is an outstanding question. The metal position depends on the preparation conditions and dramatically affects the properties of the intercalated system. The use of high temperatures in the current study ($\sim$800 C)  desorbs most of the deposited Eu and generates the cluster phase. Other phases are possible in the system for lower annealing temperatures. A similar cluster phase has also been observed for intercalated Au in graphene on SiC achieved at relatively high temperatures $\sim$700 C\cite{NarayananNair2016}. The Au cluster position is also defined by the 6$\times$6 supercell with average separation between the clusters $\sim$2.2 nm. Moreover, this study  suggests that, the Au or Eu formed clusters between the buffer layer and graphene, not only explain the preference of nucleation to be at the vertices  of the 6$\times$6 supercell but also that the cluster phase  is a more general phenomenon of metal intercalation into graphene-SiC.  

XMCD measurements are performed at the 4-ID-C beamline at the Advanced Photon Source (Argonne National Laboratory) in a chamber equipped with a high magnetic field ($<7$ T) produced by a split-coil superconducting magnet. Field dependence of the XMCD spectra are collected in helicity-switching mode in external magnetic fields applied parallel to the incident x-ray wave vector at energies that cover the Eu $M_4$  (1158 eV) and $M_5$  (1127 eV) binding energies. Measurements of x-ray absorption spectroscopy (XAS) signals are collected by total electron yield (TEY). For data analysis and normalization, the individual XAS, $\mu_+$ and $\mu_-$, are normalized by their respective monitors to compensate for incident-beam intensity variations. For the initial background subtraction, the XAS ($\mu_+$ and $\mu_-$) has a flat value subtracted such that the lowest energy (i.e. sufficiently far from the edge) is at 0 intensity, removing both background and offsets due to the beam. The total XAS ($\mu_++\mu_-$) is then scaled by a factor such that its maximum intensity is 1. That scale factor is then used to also scale the individual ($\mu_+$ and $\mu_-$) XAS. The XMCD signal is obtained from the difference between two XAS spectra of the left- and right-handed helicities, $\mu_+$  and $\mu_-$. More details on data reduction is provided elsewhere\cite{Anderson2017}. We note that our intercalated samples are removed from the ultra-high vacuum chamber and transported in air for the XMCD experiments.  As we discuss below and in the SI\cite{SI-EU}, we have also tested the samples after exposure of 9 months in air.  

\section*{Results and Discussion}

\begin{figure}\centering \includegraphics [width = 76 mm] {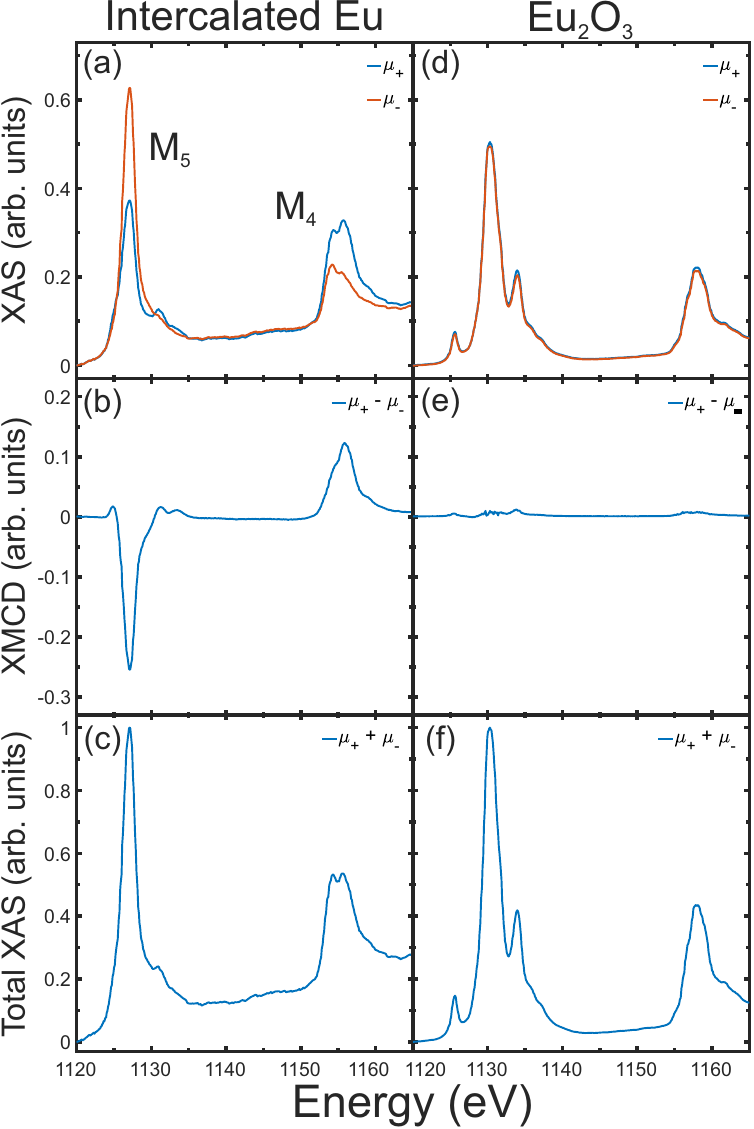}
\caption{(Color online) The XAS, XMCD, and total XAS of the intercalated Eu (left) and \ce{Eu2O3} (right) at $B=5$ T and $T=15$ K. The total XAS signals for  intercalated Eu and \ce{Eu2O3} are consistent with Eu$^{2+}$ and Eu$^{3+}$, respectively. }
\label{fig:EuXAS} 
\end{figure}

Figure\ \ref{fig:EuXAS} shows the XAS, XMCD, and total XAS at the Eu $M_4$ and $M_5$ edges at $T =15$ K and $B = 5$ T for intercalated Eu (left) and for \ce{Eu2O3} (right). We measure  \ce{Eu2O3}  as a control to monitor possible oxidation of our sample as it is exposed to air. Each of the three signals shows a significant contrast between the two samples. Figures \ref{fig:EuXAS}a and \ref{fig:EuXAS}d show the XAS of the intercalated Eu and \ce{Eu2O3} with the latter exhibiting noticeable splitting of the $M_5$ peak, which has been documented as corresponding to Eu$^{3+}$ \cite{Thole1985a,Mizumaki2005}. However, the intercalated Eu has a very prominent difference between the $\mu_+$ and $\mu_-$ while the \ce{Eu2O3} has almost none. This leads to a strong XMCD signal for the intercalated (Fig. \ref{fig:EuXAS}b) but to nearly flat XMCD signal for the oxide (Fig. \ref{fig:EuXAS}e). The zero XMCD signal for \ce{Eu2O3} is expected for the non-magnetic Eu$^{3+}$ where $L=S=3$ and a total moment $J=0$\cite{Concas2011}.

\begin{figure}\centering \includegraphics [width = 76 mm] {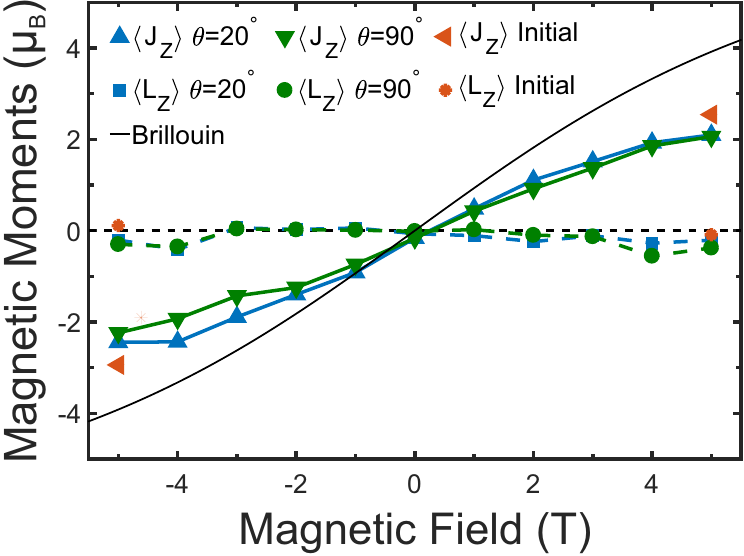}
\caption{(Color online) The magnetic field dependence of the $\JZ$ (triangles) and $\LZ$ (square and circle) of intercalated Eu at $T = 15$ K. To check for anisotropy, measurements were conducted at incident beam angles of 20$^{\circ}$ (blue) and 90$^{\circ}$ (green). The $\LZ$ components are nearly 0, which is consistent with Hund's rules for Eu$^{2+}$. The calculated Brillouin function for Eu$^{2+}$ at $T=15 K$ is also included for comparison as a smooth solid line.}
\label{fig:EuField}
\end{figure}

The XMCD of the intercalated Eu enables to quantitatively determine the orbital, $\LZ$, and spin, $\SZ$, contributions to the total moment, $\JZ$, of Eu$^{2+}$ via sum rules derived by Carra \textit{et. al}.\cite{Carra1993} as follows:
\begin{equation}
\langle L_Z \rangle=\frac{2(p+q)}{r}\nH 
\label{eq:LZ}
\end{equation}
 and 
 \begin{equation}
\SZ=\frac{2p-3q}{2r}\nH-3\TZ \approx \frac{2p-3q}{2r}\nH
\label{eq:SZ}
 \end{equation}

where $p=\int_{M_5}\mu_+-\mu_-$, $q=\int_{M_4}\mu_+-\mu_-$, $r=\int_{M_4+M_5}(\mu_++\mu_-)$, and $\nH$ is the number of electron holes in the valence shell ($\nH=7$ for Eu$^{2+}$) (it should be noted that our definition for $q$ differs from the $q$ used in Ref.\ \citen{Schumacher2014}).  In Eq. \ref{eq:SZ}, the $\TZ$ term vanishes due to the zero orbital moment.  We note that a strong spin moment, $\SZ$, and nearly zero orbital moment, $\LZ$,  are consistent with Hund's rules for Eu$^{2+}$ ($L=0 \mbox{; } S=J=7/2$)\cite{Schumacher2014,Carra1993,Crocombette1996,Wu1994} and thus $\SZ=\JZ$. Figure\ \ref{fig:EuField} shows moment calculations at $T = 15$  K as a function of magnetic field from +5 T to -5 T. Scans are conducted at both 20$^{\circ}$ and 90$^{\circ}$ angle between the magnetic field direction and the surface showing nearly paramagnetic-like behavior with no evidence of magnetic anisotropy.   The dependence of the moment on magnetic field shown in Fig. \ref{fig:EuField} is similar in shape to the Brillouin function (solid line) but  with smaller moment than that expected for paramagnetic Eu$^{2+}$. That the magnetic moment does not saturate at finite fields is another indication of no strong collective behavior of intercalated Eu clusters under graphene.
The fact that the magnetic moment $\JZ$ is well below its saturation value 7$\mu_B$, at high field and at the low temperature $T =15$ K, is puzzling.

We emphasize that the XMCD unequivocally determines the electronic configuration  of the intercalant as Eu$^{2+}$, as expected for metal Eu but also for ferromagnetic EuO.   Indeed, previous $M_4-M_5$ XAS measurements of Eu metal and EuO are almost indistinguishable due to the $d-f$ core levels, involved in the transitions, that are hardly influenced by the specific chemistry of the element\cite{Thole1985a}.  However, detailed comparison of our XMCD with that of thin films EuO indicate differences that point to the fact that the intercalated Eu is in its metallic state.  Also, the magnetic ground states of the metal and oxide are distinct at low temperatures.  Whereas  EuO is ferromagnetic at T$_C$ $\approx 67 $ K\cite{Wachter1972,Lettieri2003} with finite hysteresis\cite{Lettieri2003,Wackerlin2015}, Eu metal undergoes an incommensurate  helical magnetic structure at T$_N\approx 91 $ K\cite{Nereson1964,Jensen1991}.  As shown in Fig.\ \ref{fig:EuField} there is no evidence of magnetic moment saturation or anisotropy that is expected from a ferromagnet ruling out the possibility that the intercalated Eu is an oxide (i.e., EuO).  Another possibility is that intercalated Eu under graphene adopts the $(\sqrt{3}\times\sqrt{3})$R$30^\circ$ superstructure as an intercalated Eu in highly pyrolytic graphite (HOPG) crystals, namely, \ce{C6Eu}\cite{Suematsu1983}.   However, magnetization and specific  heat  of \ce{C6Eu}
indicate it becomes antiferromagnetic (AFM) at about 40 K. This scenario can also be discarded since AFM systems do not yield XMCD signals, and we do observe a strong XMCD signal below 40 K in our samples. 

\begin{figure}\centering \includegraphics [width = 76 mm] {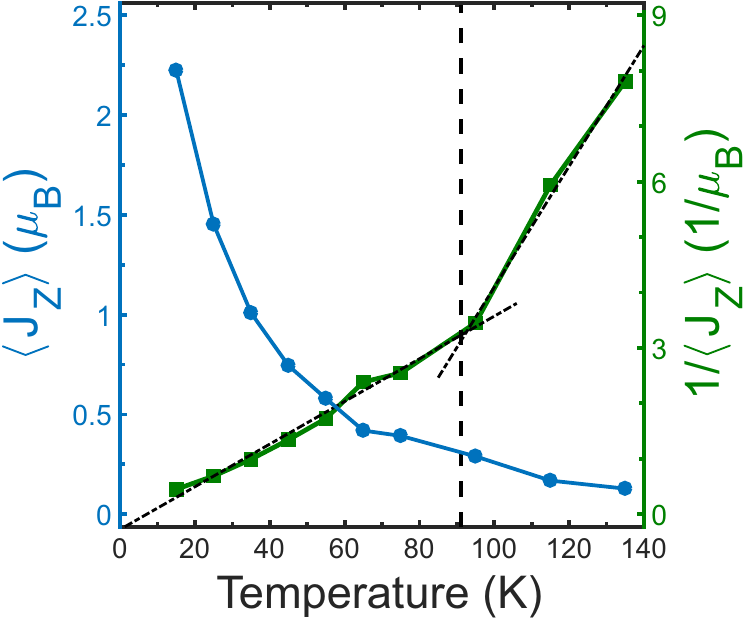}
\caption{(Color online) Temperature dependence of the total moment $\JZ$ and 1/$\JZ$ for Intercalated Eu at $B = 5$ T. Bulk Eu has a transition to helical structure at 91 K, which is indicated by the vertical dashed line. The two dashed lines are linear fits below and above two temperature regions with intersection at $\approx 90$ K.}
\label{fig:EuTemp} 
\end{figure}

To further explore the magnetic properties of the intercalated Eu nano-clusters, we have collected XMCD spectra at the $M_5$  regime (from 1120 to 1140 eV) at various temperatures and at fixed  $B = 5$ T.  As discussed in the SI\cite{SI-EU}, because  $\LZ =0$ for Eu$^{2+}$, measuring the XMCD on either the M$_5$ or M$_4$ is sufficient to determine the magnetic moment.  Figure \ref{fig:EuTemp} shows the temperature dependence of $\JZ$ from the XMCD spectra for the M$_5$ as a function of temperature, with characteristic increase common to a paramagnetic system.  However, the 1/$\JZ$ of the same data shows two distinct regions that  overlaid by linear fits (dashed lines)  intersect at $T^* \approx 90 $ K. We note that $T^*$ is very close to the  the N{\`e}el temperature, $T_N$, of bulk metallic Eu at 91 K (vertical dashed line in Fig.\ \ref{fig:EuTemp})\cite{Nereson1964,Jensen1991}. As mentioned previously, the Curie temperature of EuO is at $T_c \approx 67$ K, which is substantially lower than the anomaly  observed in our temperature dependence\cite{Wachter1972,Lettieri2003}.  This is yet another indication that the intercalated Eu-clusters under graphene are likely in their metallic structure.  In the SI we propose three scenarios of possible layers underneath graphene that may also explain the finite clustering size $2.5$ nm in diameter.     

\section*{Conclusions}
In conclusion, we have succeeded to intercalate Eu under epitaxial graphene on SiC buffer layer. Our XMCD results show the electronic configuration of the intercalant is  that of Eu$^{2+}$ likely in its metallic state or as a Eu-silicide\cite{Averyanov2016}. Our STM images show that the Eu forms relatively uniform nano-clusters of approximately 2.5 nm in diameter, and although the clusters are randomly distributed they preferably nucleate at the vertices of the 6$\times$6 super structure of graphene on SiC which act as nucleation centers.  We argue that unlike intercalated \ce{C6Eu}, the Eu under graphene forms clusters that  likely conform to the square unit cell of metallic Eu and that, due to the incommensurabilty between graphene and the Eu, the  clusters are limited in size. The temperature dependence of $\JZ$ at fixed magnetic field $B=5$ T is consistent with the paramagnetic behavior displayed in the magnetic field dependence at $T=15$ K, namely, no anisotropy or hysteresis effects are observed.  Although Eu  is a highly oxidizing metal in air, the epitaxial graphene layer formed on SiC is practically defect free that protects the intercalated Eu against oxidation under atmospheric conditions over periods of months.

\section{Acknowledgments}
Ames Laboratory is operated by Iowa State University by support from the U.S. Department of Energy, Office of Basic Energy Sciences, under Contract No. DE-AC02-07CH11358. Use of the Advanced Photon Source, an Office of Science User Facility operated for the U.S. Department of Energy (DOE) Office of Science by Argonne National Laboratory, is supported by the U.S. DOE under Contract No. DE-AC02-06CH11357.

\bibliographystyle{apsrev4-2}

\bibliography{IntercalatedEu.bbl}

\begin{thebibliography}{33}%
\makeatletter
\providecommand \@ifxundefined [1]{%
 \@ifx{#1\undefined}
}%
\providecommand \@ifnum [1]{%
 \ifnum #1\expandafter \@firstoftwo
 \else \expandafter \@secondoftwo
 \fi
}%
\providecommand \@ifx [1]{%
 \ifx #1\expandafter \@firstoftwo
 \else \expandafter \@secondoftwo
 \fi
}%
\providecommand \natexlab [1]{#1}%
\providecommand \enquote  [1]{``#1''}%
\providecommand \bibnamefont  [1]{#1}%
\providecommand \bibfnamefont [1]{#1}%
\providecommand \citenamefont [1]{#1}%
\providecommand \href@noop [0]{\@secondoftwo}%
\providecommand \href [0]{\begingroup \@sanitize@url \@href}%
\providecommand \@href[1]{\@@startlink{#1}\@@href}%
\providecommand \@@href[1]{\endgroup#1\@@endlink}%
\providecommand \@sanitize@url [0]{\catcode `\\12\catcode `\$12\catcode
  `\&12\catcode `\#12\catcode `\^12\catcode `\_12\catcode `\%12\relax}%
\providecommand \@@startlink[1]{}%
\providecommand \@@endlink[0]{}%
\providecommand \url  [0]{\begingroup\@sanitize@url \@url }%
\providecommand \@url [1]{\endgroup\@href {#1}{\urlprefix }}%
\providecommand \urlprefix  [0]{URL }%
\providecommand \Eprint [0]{\href }%
\providecommand \doibase [0]{http://dx.doi.org/}%
\providecommand \selectlanguage [0]{\@gobble}%
\providecommand \bibinfo  [0]{\@secondoftwo}%
\providecommand \bibfield  [0]{\@secondoftwo}%
\providecommand \translation [1]{[#1]}%
\providecommand \BibitemOpen [0]{}%
\providecommand \bibitemStop [0]{}%
\providecommand \bibitemNoStop [0]{.\EOS\space}%
\providecommand \EOS [0]{\spacefactor3000\relax}%
\providecommand \BibitemShut  [1]{\csname bibitem#1\endcsname}%
\let\auto@bib@innerbib\@empty
\bibitem [{\citenamefont {Zhang}\ \emph {et~al.}(2014)\citenamefont {Zhang},
  \citenamefont {Rajaraman}, \citenamefont {Liu},\ and\ \citenamefont
  {Ramakrishna}}]{Zhang2014a}%
  \BibitemOpen
  \bibfield  {author} {\bibinfo {author} {\bibfnamefont {X.}~\bibnamefont
  {Zhang}}, \bibinfo {author} {\bibfnamefont {B.~R.~S.}\ \bibnamefont
  {Rajaraman}}, \bibinfo {author} {\bibfnamefont {H.}~\bibnamefont {Liu}}, \
  and\ \bibinfo {author} {\bibfnamefont {S.}~\bibnamefont {Ramakrishna}},\
  }\bibfield  {title} {\emph {\enquote {\bibinfo {title} {Graphene's potential
  in materials science and engineering},}\ }}\href {\doibase
  10.1039/C4RA02817A} {\bibfield  {journal} {\bibinfo  {journal} {RSC Adv.}\
  }\textbf {\bibinfo {volume} {4}},\ \bibinfo {pages} {28987} (\bibinfo {year}
  {2014})}\BibitemShut {NoStop}%
\bibitem [{\citenamefont {Dedkov}\ \emph
  {et~al.}(2008{\natexlab{a}})\citenamefont {Dedkov}, \citenamefont {Fonin},\
  and\ \citenamefont {Laubschat}}]{Dedkov2008a}%
  \BibitemOpen
  \bibfield  {author} {\bibinfo {author} {\bibfnamefont {Y.~S.}\ \bibnamefont
  {Dedkov}}, \bibinfo {author} {\bibfnamefont {M.}~\bibnamefont {Fonin}}, \
  and\ \bibinfo {author} {\bibfnamefont {C.}~\bibnamefont {Laubschat}},\
  }\bibfield  {title} {\emph {\enquote {\bibinfo {title} {A possible source of
  spin-polarized electrons: {{The}} inert graphene/{{Ni}}(111) system},}\
  }}\href {\doibase 10.1063/1.2841809} {\bibfield  {journal} {\bibinfo
  {journal} {Appl. Phys. Lett.}\ }\textbf {\bibinfo {volume} {92}},\ \bibinfo
  {pages} {052506} (\bibinfo {year} {2008}{\natexlab{a}})}\BibitemShut
  {NoStop}%
\bibitem [{\citenamefont {Dedkov}\ \emph
  {et~al.}(2008{\natexlab{b}})\citenamefont {Dedkov}, \citenamefont {Fonin},
  \citenamefont {R{\"u}diger},\ and\ \citenamefont {Laubschat}}]{Dedkov2008}%
  \BibitemOpen
  \bibfield  {author} {\bibinfo {author} {\bibfnamefont {Y.~S.}\ \bibnamefont
  {Dedkov}}, \bibinfo {author} {\bibfnamefont {M.}~\bibnamefont {Fonin}},
  \bibinfo {author} {\bibfnamefont {U.}~\bibnamefont {R{\"u}diger}}, \ and\
  \bibinfo {author} {\bibfnamefont {C.}~\bibnamefont {Laubschat}},\ }\bibfield
  {title} {\emph {\enquote {\bibinfo {title} {Graphene-protected iron layer on
  {{Ni}}(111)},}\ }}\href {\doibase 10.1063/1.2953972} {\bibfield  {journal}
  {\bibinfo  {journal} {Appl. Phys. Lett.}\ }\textbf {\bibinfo {volume} {93}},\
  \bibinfo {pages} {022509} (\bibinfo {year} {2008}{\natexlab{b}})}\BibitemShut
  {NoStop}%
\bibitem [{\citenamefont {B{\"o}hm}(2014)}]{Bohm2014}%
  \BibitemOpen
  \bibfield  {author} {\bibinfo {author} {\bibfnamefont {S.}~\bibnamefont
  {B{\"o}hm}},\ }\bibfield  {title} {\emph {\enquote {\bibinfo {title}
  {Graphene against corrosion},}\ }}\href {\doibase 10.1038/nnano.2014.220}
  {\bibfield  {journal} {\bibinfo  {journal} {Nat. Nanotech.}\ }\textbf
  {\bibinfo {volume} {9}},\ \bibinfo {pages} {741} (\bibinfo {year}
  {2014})}\BibitemShut {NoStop}%
\bibitem [{\citenamefont {Bunch}\ \emph {et~al.}(2008)\citenamefont {Bunch},
  \citenamefont {Verbridge}, \citenamefont {Alden}, \citenamefont {{van der
  Zande}}, \citenamefont {Parpia}, \citenamefont {Craighead},\ and\
  \citenamefont {McEuen}}]{Bunch2008}%
  \BibitemOpen
  \bibfield  {author} {\bibinfo {author} {\bibfnamefont {J.~S.}\ \bibnamefont
  {Bunch}}, \bibinfo {author} {\bibfnamefont {S.~S.}\ \bibnamefont
  {Verbridge}}, \bibinfo {author} {\bibfnamefont {J.~S.}\ \bibnamefont
  {Alden}}, \bibinfo {author} {\bibfnamefont {A.~M.}\ \bibnamefont {{van der
  Zande}}}, \bibinfo {author} {\bibfnamefont {J.~M.}\ \bibnamefont {Parpia}},
  \bibinfo {author} {\bibfnamefont {H.~G.}\ \bibnamefont {Craighead}}, \ and\
  \bibinfo {author} {\bibfnamefont {P.~L.}\ \bibnamefont {McEuen}},\ }\bibfield
   {title} {\emph {\enquote {\bibinfo {title} {Impermeable {{Atomic Membranes}}
  from {{Graphene Sheets}}},}\ }}\href {\doibase 10.1021/nl801457b} {\bibfield
  {journal} {\bibinfo  {journal} {Nano Lett.}\ }\textbf {\bibinfo {volume}
  {8}},\ \bibinfo {pages} {2458} (\bibinfo {year} {2008})}\BibitemShut
  {NoStop}%
\bibitem [{\citenamefont {Hofmann}\ \emph {et~al.}(2015)\citenamefont
  {Hofmann}, \citenamefont {Braeuninger-Weimer},\ and\ \citenamefont
  {Weatherup}}]{Hofmann2015}%
  \BibitemOpen
  \bibfield  {author} {\bibinfo {author} {\bibfnamefont {S.}~\bibnamefont
  {Hofmann}}, \bibinfo {author} {\bibfnamefont {P.}~\bibnamefont
  {Braeuninger-Weimer}}, \ and\ \bibinfo {author} {\bibfnamefont {R.~S.}\
  \bibnamefont {Weatherup}},\ }\bibfield  {title} {\emph {\enquote {\bibinfo
  {title} {{{CVD}}-{{Enabled Graphene Manufacture}} and {{Technology}}},}\
  }}\href {\doibase 10.1021/acs.jpclett.5b01052} {\bibfield  {journal}
  {\bibinfo  {journal} {J. Phys. Chem. Lett.}\ }\textbf {\bibinfo {volume}
  {6}},\ \bibinfo {pages} {2714} (\bibinfo {year} {2015})}\BibitemShut
  {NoStop}%
\bibitem [{\citenamefont {Weatherup}\ \emph {et~al.}(2015)\citenamefont
  {Weatherup}, \citenamefont {D'Arsi{\'e}}, \citenamefont {Cabrero-Vilatela},
  \citenamefont {Caneva}, \citenamefont {Blume}, \citenamefont {Robertson},
  \citenamefont {Schloegl},\ and\ \citenamefont {Hofmann}}]{Weatherup2015}%
  \BibitemOpen
  \bibfield  {author} {\bibinfo {author} {\bibfnamefont {R.~S.}\ \bibnamefont
  {Weatherup}}, \bibinfo {author} {\bibfnamefont {L.}~\bibnamefont
  {D'Arsi{\'e}}}, \bibinfo {author} {\bibfnamefont {A.}~\bibnamefont
  {Cabrero-Vilatela}}, \bibinfo {author} {\bibfnamefont {S.}~\bibnamefont
  {Caneva}}, \bibinfo {author} {\bibfnamefont {R.}~\bibnamefont {Blume}},
  \bibinfo {author} {\bibfnamefont {J.}~\bibnamefont {Robertson}}, \bibinfo
  {author} {\bibfnamefont {R.}~\bibnamefont {Schloegl}}, \ and\ \bibinfo
  {author} {\bibfnamefont {S.}~\bibnamefont {Hofmann}},\ }\bibfield  {title}
  {\emph {\enquote {\bibinfo {title} {Long-{{Term Passivation}} of {{Strongly
  Interacting Metals}} with {{Single}}-{{Layer Graphene}}},}\ }}\href {\doibase
  10.1021/jacs.5b08729} {\bibfield  {journal} {\bibinfo  {journal} {J. Am.
  Chem. Soc.}\ }\textbf {\bibinfo {volume} {137}},\ \bibinfo {pages} {14358}
  (\bibinfo {year} {2015})}\BibitemShut {NoStop}%
\bibitem [{\citenamefont {Anderson}\ \emph
  {et~al.}(2017{\natexlab{a}})\citenamefont {Anderson}, \citenamefont {Zhang},
  \citenamefont {Hupalo}, \citenamefont {Rosenberg}, \citenamefont
  {Tringides},\ and\ \citenamefont {Vaknin}}]{Anderson2017a}%
  \BibitemOpen
  \bibfield  {author} {\bibinfo {author} {\bibfnamefont {N.~A.}\ \bibnamefont
  {Anderson}}, \bibinfo {author} {\bibfnamefont {Q.}~\bibnamefont {Zhang}},
  \bibinfo {author} {\bibfnamefont {M.}~\bibnamefont {Hupalo}}, \bibinfo
  {author} {\bibfnamefont {R.~A.}\ \bibnamefont {Rosenberg}}, \bibinfo {author}
  {\bibfnamefont {M.~C.}\ \bibnamefont {Tringides}}, \ and\ \bibinfo {author}
  {\bibfnamefont {D.}~\bibnamefont {Vaknin}},\ }\bibfield  {title} {\emph
  {\enquote {\bibinfo {title} {Magnetite nano-islands on silicon-carbide with
  graphene},}\ }}\href {\doibase 10.1063/1.4973571} {\bibfield  {journal}
  {\bibinfo  {journal} {J. Appl. Phys.}\ }\textbf {\bibinfo {volume} {121}},\
  \bibinfo {pages} {014310} (\bibinfo {year} {2017}{\natexlab{a}})}\BibitemShut
  {NoStop}%
\bibitem [{\citenamefont {Forbeaux}\ \emph {et~al.}(1998)\citenamefont
  {Forbeaux}, \citenamefont {Themlin},\ and\ \citenamefont
  {Debever}}]{Forbeaux1998}%
  \BibitemOpen
  \bibfield  {author} {\bibinfo {author} {\bibfnamefont {I.}~\bibnamefont
  {Forbeaux}}, \bibinfo {author} {\bibfnamefont {J.-M.}\ \bibnamefont
  {Themlin}}, \ and\ \bibinfo {author} {\bibfnamefont {J.-M.}\ \bibnamefont
  {Debever}},\ }\bibfield  {title} {\emph {\enquote {\bibinfo {title}
  {Heteroepitaxial graphite on {{6H}}-{{SiC}}(0001): {{Interface}} formation
  through conduction-band electronic structure},}\ }}\href {\doibase
  10.1103/PhysRevB.58.16396} {\bibfield  {journal} {\bibinfo  {journal} {Phys.
  Rev. B}\ }\textbf {\bibinfo {volume} {58}},\ \bibinfo {pages} {16396}
  (\bibinfo {year} {1998})}\BibitemShut {NoStop}%
\bibitem [{\citenamefont {Ren}\ \emph {et~al.}(2010)\citenamefont {Ren},
  \citenamefont {Chen}, \citenamefont {Cai}, \citenamefont {Zhu}, \citenamefont
  {Zhu},\ and\ \citenamefont {Ruoff}}]{ren2010}%
  \BibitemOpen
  \bibfield  {author} {\bibinfo {author} {\bibfnamefont {Y.}~\bibnamefont
  {Ren}}, \bibinfo {author} {\bibfnamefont {S.}~\bibnamefont {Chen}}, \bibinfo
  {author} {\bibfnamefont {W.}~\bibnamefont {Cai}}, \bibinfo {author}
  {\bibfnamefont {Y.}~\bibnamefont {Zhu}}, \bibinfo {author} {\bibfnamefont
  {C.}~\bibnamefont {Zhu}}, \ and\ \bibinfo {author} {\bibfnamefont {R.~S.}\
  \bibnamefont {Ruoff}},\ }\bibfield  {title} {\emph {\enquote {\bibinfo
  {title} {Controlling the electrical transport properties of graphene by
  {\emph{in situ}} metal deposition},}\ }}\href {\doibase 10.1063/1.3471396}
  {\bibfield  {journal} {\bibinfo  {journal} {Appl. Phys. Lett.}\ }\textbf
  {\bibinfo {volume} {97}},\ \bibinfo {pages} {053107} (\bibinfo {year}
  {2010})}\BibitemShut {NoStop}%
\bibitem [{\citenamefont {F{\"o}rster}\ \emph {et~al.}(2012)\citenamefont
  {F{\"o}rster}, \citenamefont {Wehling}, \citenamefont {Schumacher},
  \citenamefont {Rosch},\ and\ \citenamefont {Michely}}]{Forster2012}%
  \BibitemOpen
  \bibfield  {author} {\bibinfo {author} {\bibfnamefont {D.~F.}\ \bibnamefont
  {F{\"o}rster}}, \bibinfo {author} {\bibfnamefont {T.~O.}\ \bibnamefont
  {Wehling}}, \bibinfo {author} {\bibfnamefont {S.}~\bibnamefont {Schumacher}},
  \bibinfo {author} {\bibfnamefont {A.}~\bibnamefont {Rosch}}, \ and\ \bibinfo
  {author} {\bibfnamefont {T.}~\bibnamefont {Michely}},\ }\bibfield  {title}
  {\emph {\enquote {\bibinfo {title} {Phase coexistence of clusters and
  islands: Europium on graphene},}\ }}\href {\doibase
  10.1088/1367-2630/14/2/023022} {\bibfield  {journal} {\bibinfo  {journal}
  {New J. Phys.}\ }\textbf {\bibinfo {volume} {14}},\ \bibinfo {pages} {023022}
  (\bibinfo {year} {2012})}\BibitemShut {NoStop}%
\bibitem [{\citenamefont {Wang}\ \emph {et~al.}(2008)\citenamefont {Wang},
  \citenamefont {Tabakman},\ and\ \citenamefont {Dai}}]{Wang2008}%
  \BibitemOpen
  \bibfield  {author} {\bibinfo {author} {\bibfnamefont {X.}~\bibnamefont
  {Wang}}, \bibinfo {author} {\bibfnamefont {S.~M.}\ \bibnamefont {Tabakman}},
  \ and\ \bibinfo {author} {\bibfnamefont {H.}~\bibnamefont {Dai}},\ }\bibfield
   {title} {\emph {\enquote {\bibinfo {title} {Atomic {{Layer Deposition}} of
  {{Metal Oxides}} on {{Pristine}} and {{Functionalized Graphene}}},}\ }}\href
  {\doibase 10.1021/ja8023059} {\bibfield  {journal} {\bibinfo  {journal} {J.
  Am. Chem. Soc.}\ }\textbf {\bibinfo {volume} {130}},\ \bibinfo {pages} {8152}
  (\bibinfo {year} {2008})}\BibitemShut {NoStop}%
\bibitem [{\citenamefont {Premlal}\ \emph {et~al.}(2009)\citenamefont
  {Premlal}, \citenamefont {Cranney}, \citenamefont {Vonau}, \citenamefont
  {Aubel}, \citenamefont {Casterman}, \citenamefont {Souza},\ and\
  \citenamefont {Simon}}]{Premlal2009}%
  \BibitemOpen
  \bibfield  {author} {\bibinfo {author} {\bibfnamefont {B.}~\bibnamefont
  {Premlal}}, \bibinfo {author} {\bibfnamefont {M.}~\bibnamefont {Cranney}},
  \bibinfo {author} {\bibfnamefont {F.}~\bibnamefont {Vonau}}, \bibinfo
  {author} {\bibfnamefont {D.}~\bibnamefont {Aubel}}, \bibinfo {author}
  {\bibfnamefont {D.}~\bibnamefont {Casterman}}, \bibinfo {author}
  {\bibfnamefont {M.~M.~D.}\ \bibnamefont {Souza}}, \ and\ \bibinfo {author}
  {\bibfnamefont {L.}~\bibnamefont {Simon}},\ }\bibfield  {title} {\emph
  {\enquote {\bibinfo {title} {Surface intercalation of gold underneath a
  graphene monolayer on {{SiC}}(0001) studied by scanning tunneling microscopy
  and spectroscopy},}\ }}\href {\doibase 10.1063/1.3168502} {\bibfield
  {journal} {\bibinfo  {journal} {Appl. Phys. Lett.}\ }\textbf {\bibinfo
  {volume} {94}},\ \bibinfo {pages} {263115} (\bibinfo {year}
  {2009})}\BibitemShut {NoStop}%
\bibitem [{\citenamefont {Schumacher}\ \emph {et~al.}(2014)\citenamefont
  {Schumacher}, \citenamefont {Huttmann}, \citenamefont {Petrovi{\'c}},
  \citenamefont {Witt}, \citenamefont {F{\"o}rster}, \citenamefont {Vo-Van},
  \citenamefont {Coraux}, \citenamefont {Mart{\'\i}nez-Galera}, \citenamefont
  {Sessi}, \citenamefont {Vergara}, \citenamefont {R{\"u}ckamp}, \citenamefont
  {Gr{\"u}ninger}, \citenamefont {Schleheck}, \citenamefont {{Meyer zu
  Heringdorf}}, \citenamefont {Ohresser}, \citenamefont {Kralj}, \citenamefont
  {Wehling},\ and\ \citenamefont {Michely}}]{Schumacher2014}%
  \BibitemOpen
  \bibfield  {author} {\bibinfo {author} {\bibfnamefont {S.}~\bibnamefont
  {Schumacher}}, \bibinfo {author} {\bibfnamefont {F.}~\bibnamefont
  {Huttmann}}, \bibinfo {author} {\bibfnamefont {M.}~\bibnamefont
  {Petrovi{\'c}}}, \bibinfo {author} {\bibfnamefont {C.}~\bibnamefont {Witt}},
  \bibinfo {author} {\bibfnamefont {D.~F.}\ \bibnamefont {F{\"o}rster}},
  \bibinfo {author} {\bibfnamefont {C.}~\bibnamefont {Vo-Van}}, \bibinfo
  {author} {\bibfnamefont {J.}~\bibnamefont {Coraux}}, \bibinfo {author}
  {\bibfnamefont {A.~J.}\ \bibnamefont {Mart{\'\i}nez-Galera}}, \bibinfo
  {author} {\bibfnamefont {V.}~\bibnamefont {Sessi}}, \bibinfo {author}
  {\bibfnamefont {I.}~\bibnamefont {Vergara}}, \bibinfo {author} {\bibfnamefont
  {R.}~\bibnamefont {R{\"u}ckamp}}, \bibinfo {author} {\bibfnamefont
  {M.}~\bibnamefont {Gr{\"u}ninger}}, \bibinfo {author} {\bibfnamefont
  {N.}~\bibnamefont {Schleheck}}, \bibinfo {author} {\bibfnamefont
  {F.}~\bibnamefont {{Meyer zu Heringdorf}}}, \bibinfo {author} {\bibfnamefont
  {P.}~\bibnamefont {Ohresser}}, \bibinfo {author} {\bibfnamefont
  {M.}~\bibnamefont {Kralj}}, \bibinfo {author} {\bibfnamefont {T.~O.}\
  \bibnamefont {Wehling}}, \ and\ \bibinfo {author} {\bibfnamefont
  {T.}~\bibnamefont {Michely}},\ }\bibfield  {title} {\emph {\enquote {\bibinfo
  {title} {Europium underneath graphene on {{Ir}}(111): {{Intercalation}}
  mechanism, magnetism, and band structure},}\ }}\href {\doibase
  10.1103/PhysRevB.90.235437} {\bibfield  {journal} {\bibinfo  {journal} {Phys.
  Rev. B}\ }\textbf {\bibinfo {volume} {90}},\ \bibinfo {pages} {235437}
  (\bibinfo {year} {2014})}\BibitemShut {NoStop}%
\bibitem [{\citenamefont {Voloshina}\ and\ \citenamefont
  {Dedkov}(2014)}]{Voloshina2014}%
  \BibitemOpen
  \bibfield  {author} {\bibinfo {author} {\bibfnamefont {E.~N.}\ \bibnamefont
  {Voloshina}}\ and\ \bibinfo {author} {\bibfnamefont {Y.~S.}\ \bibnamefont
  {Dedkov}},\ }\bibfield  {title} {\emph {\enquote {\bibinfo {title}
  {Electronic and magnetic properties of the graphene/{{Eu}}/{{Ni}}(111) hybrid
  system},}\ }}\href {\doibase 10.5560/ZNA.2014-0012} {\bibfield  {journal}
  {\bibinfo  {journal} {Z. Naturforsch. A}\ }\textbf {\bibinfo {volume} {69}},\
  \bibinfo {pages} {297} (\bibinfo {year} {2014})},\ \Eprint
  {http://arxiv.org/abs/1407.4389} {arXiv:1407.4389} \BibitemShut {NoStop}%
\bibitem [{\citenamefont {Huttmann}\ \emph {et~al.}(2017)\citenamefont
  {Huttmann}, \citenamefont {Klar}, \citenamefont {Atodiresei}, \citenamefont
  {Schmitz-Antoniak}, \citenamefont {Smekhova}, \citenamefont
  {Mart{\'\i}nez-Galera}, \citenamefont {Caciuc}, \citenamefont {Bihlmayer},
  \citenamefont {Bl{\"u}gel}, \citenamefont {Michely},\ and\ \citenamefont
  {Wende}}]{Huttmann2017}%
  \BibitemOpen
  \bibfield  {author} {\bibinfo {author} {\bibfnamefont {F.}~\bibnamefont
  {Huttmann}}, \bibinfo {author} {\bibfnamefont {D.}~\bibnamefont {Klar}},
  \bibinfo {author} {\bibfnamefont {N.}~\bibnamefont {Atodiresei}}, \bibinfo
  {author} {\bibfnamefont {C.}~\bibnamefont {Schmitz-Antoniak}}, \bibinfo
  {author} {\bibfnamefont {A.}~\bibnamefont {Smekhova}}, \bibinfo {author}
  {\bibfnamefont {A.~J.}\ \bibnamefont {Mart{\'\i}nez-Galera}}, \bibinfo
  {author} {\bibfnamefont {V.}~\bibnamefont {Caciuc}}, \bibinfo {author}
  {\bibfnamefont {G.}~\bibnamefont {Bihlmayer}}, \bibinfo {author}
  {\bibfnamefont {S.}~\bibnamefont {Bl{\"u}gel}}, \bibinfo {author}
  {\bibfnamefont {T.}~\bibnamefont {Michely}}, \ and\ \bibinfo {author}
  {\bibfnamefont {H.}~\bibnamefont {Wende}},\ }\bibfield  {title} {\emph
  {\enquote {\bibinfo {title} {Magnetism in a graphene-4f-3d hybrid system},}\
  }}\href {\doibase 10.1103/PhysRevB.95.075427} {\bibfield  {journal} {\bibinfo
   {journal} {Phys. Rev. B}\ }\textbf {\bibinfo {volume} {95}},\ \bibinfo
  {pages} {075427} (\bibinfo {year} {2017})}\BibitemShut {NoStop}%
\bibitem [{\citenamefont {Sung}\ \emph {et~al.}(2017)\citenamefont {Sung},
  \citenamefont {Kim}, \citenamefont {Lee}, \citenamefont {Kim}, \citenamefont
  {Ryu}, \citenamefont {Park}, \citenamefont {Kim}, \citenamefont {Min},\ and\
  \citenamefont {Chung}}]{Sung2017}%
  \BibitemOpen
  \bibfield  {author} {\bibinfo {author} {\bibfnamefont {S.}~\bibnamefont
  {Sung}}, \bibinfo {author} {\bibfnamefont {S.}~\bibnamefont {Kim}}, \bibinfo
  {author} {\bibfnamefont {P.}~\bibnamefont {Lee}}, \bibinfo {author}
  {\bibfnamefont {J.}~\bibnamefont {Kim}}, \bibinfo {author} {\bibfnamefont
  {M.}~\bibnamefont {Ryu}}, \bibinfo {author} {\bibfnamefont {H.}~\bibnamefont
  {Park}}, \bibinfo {author} {\bibfnamefont {K.}~\bibnamefont {Kim}}, \bibinfo
  {author} {\bibfnamefont {B.~I.}\ \bibnamefont {Min}}, \ and\ \bibinfo
  {author} {\bibfnamefont {J.}~\bibnamefont {Chung}},\ }\bibfield  {title}
  {\emph {\enquote {\bibinfo {title} {Observation of variable hybridized-band
  gaps in {{Eu}}-intercalated graphene},}\ }}\href {\doibase
  10.1088/1361-6528/aa6951} {\bibfield  {journal} {\bibinfo  {journal}
  {Nanotechnology}\ }\textbf {\bibinfo {volume} {28}},\ \bibinfo {pages}
  {205201} (\bibinfo {year} {2017})}\BibitemShut {NoStop}%
\bibitem [{\citenamefont {Dresselhaus}\ and\ \citenamefont
  {Dresselhaus}(1981)}]{Dresselhaus1981}%
  \BibitemOpen
  \bibfield  {author} {\bibinfo {author} {\bibfnamefont {M.~S.}\ \bibnamefont
  {Dresselhaus}}\ and\ \bibinfo {author} {\bibfnamefont {G.}~\bibnamefont
  {Dresselhaus}},\ }\bibfield  {title} {\emph {\enquote {\bibinfo {title}
  {Intercalation compounds of graphite},}\ }}\href {\doibase
  10.1080/00018738100101367} {\bibfield  {journal} {\bibinfo  {journal} {Adv.
  Phys.}\ }\textbf {\bibinfo {volume} {30}},\ \bibinfo {pages} {139} (\bibinfo
  {year} {1981})}\BibitemShut {NoStop}%
  \bibitem{SI-EU}  See Supplementary Material at (link) for more information on the sample preparation and details on the intercalation, sample exposure to air over time as measured by XMCD, details on the calculation of the magnetic moment from the XMCD, more details on the Eu clusters under graphene and the relationship to graphite intercalation compounds, and examination of various hypotheses regarding metal-Eu plane mismatch with graphene as a mechanism to rationalize nucleation and finite average size of the Eu clusters under graphene.  
\bibitem [{\citenamefont {Narayanan~Nair}\ \emph {et~al.}(2016)\citenamefont
  {Narayanan~Nair}, \citenamefont {Cranney}, \citenamefont {Jiang},
  \citenamefont {Hajjar-Garreau}, \citenamefont {Aubel}, \citenamefont {Vonau},
  \citenamefont {Florentin}, \citenamefont {Denys}, \citenamefont {Bocquet},\
  and\ \citenamefont {Simon}}]{NarayananNair2016}%
  \BibitemOpen
  \bibfield  {author} {\bibinfo {author} {\bibfnamefont {M.}~\bibnamefont
  {Narayanan~Nair}}, \bibinfo {author} {\bibfnamefont {M.}~\bibnamefont
  {Cranney}}, \bibinfo {author} {\bibfnamefont {T.}~\bibnamefont {Jiang}},
  \bibinfo {author} {\bibfnamefont {S.}~\bibnamefont {Hajjar-Garreau}},
  \bibinfo {author} {\bibfnamefont {D.}~\bibnamefont {Aubel}}, \bibinfo
  {author} {\bibfnamefont {F.}~\bibnamefont {Vonau}}, \bibinfo {author}
  {\bibfnamefont {A.}~\bibnamefont {Florentin}}, \bibinfo {author}
  {\bibfnamefont {E.}~\bibnamefont {Denys}}, \bibinfo {author} {\bibfnamefont
  {M.-L.}\ \bibnamefont {Bocquet}}, \ and\ \bibinfo {author} {\bibfnamefont
  {L.}~\bibnamefont {Simon}},\ }\bibfield  {title} {\emph {\enquote {\bibinfo
  {title} {Noble-metal intercalation process leading to a protected adatom in a
  graphene hollow site},}\ }}\href {\doibase 10.1103/PhysRevB.94.075427}
  {\bibfield  {journal} {\bibinfo  {journal} {Phys. Rev. B}\ }\textbf {\bibinfo
  {volume} {94}},\ \bibinfo {pages} {075427} (\bibinfo {year}
  {2016})}\BibitemShut {NoStop}%
\bibitem [{\citenamefont {Anderson}\ \emph
  {et~al.}(2017{\natexlab{b}})\citenamefont {Anderson}, \citenamefont {Zhang},
  \citenamefont {Hupalo}, \citenamefont {Rosenberg}, \citenamefont {Freeland},
  \citenamefont {Tringides},\ and\ \citenamefont {Vaknin}}]{Anderson2017}%
  \BibitemOpen
  \bibfield  {author} {\bibinfo {author} {\bibfnamefont {N.~A.}\ \bibnamefont
  {Anderson}}, \bibinfo {author} {\bibfnamefont {Q.}~\bibnamefont {Zhang}},
  \bibinfo {author} {\bibfnamefont {M.}~\bibnamefont {Hupalo}}, \bibinfo
  {author} {\bibfnamefont {R.~A.}\ \bibnamefont {Rosenberg}}, \bibinfo {author}
  {\bibfnamefont {J.~W.}\ \bibnamefont {Freeland}}, \bibinfo {author}
  {\bibfnamefont {M.~C.}\ \bibnamefont {Tringides}}, \ and\ \bibinfo {author}
  {\bibfnamefont {D.}~\bibnamefont {Vaknin}},\ }\bibfield  {title} {\emph
  {\enquote {\bibinfo {title} {Magnetic properties of {{Dy}} nano-islands on
  graphene},}\ }}\href {\doibase 10.1016/j.jmmm.2017.04.007} {\bibfield
  {journal} {\bibinfo  {journal} {J. Magn. Magn. Mater.}\ }\textbf {\bibinfo
  {volume} {435}},\ \bibinfo {pages} {212} (\bibinfo {year}
  {2017}{\natexlab{b}})}\BibitemShut {NoStop}%
\bibitem [{\citenamefont {Thole}\ \emph {et~al.}(1985)\citenamefont {Thole},
  \citenamefont {{van der Laan}}, \citenamefont {Fuggle}, \citenamefont
  {Sawatzky}, \citenamefont {Karnatak},\ and\ \citenamefont
  {Esteva}}]{Thole1985a}%
  \BibitemOpen
  \bibfield  {author} {\bibinfo {author} {\bibfnamefont {B.~T.}\ \bibnamefont
  {Thole}}, \bibinfo {author} {\bibfnamefont {G.}~\bibnamefont {{van der
  Laan}}}, \bibinfo {author} {\bibfnamefont {J.~C.}\ \bibnamefont {Fuggle}},
  \bibinfo {author} {\bibfnamefont {G.~A.}\ \bibnamefont {Sawatzky}}, \bibinfo
  {author} {\bibfnamefont {R.~C.}\ \bibnamefont {Karnatak}}, \ and\ \bibinfo
  {author} {\bibfnamefont {J.-M.}\ \bibnamefont {Esteva}},\ }\bibfield  {title}
  {\emph {\enquote {\bibinfo {title} {3d x-ray-absorption lines and the
  3d{\textsuperscript{9}}4f{\textsuperscript{n+1}} multiplets of the
  lanthanides},}\ }}\href {\doibase 10.1103/PhysRevB.32.5107} {\bibfield
  {journal} {\bibinfo  {journal} {Phys. Rev. B}\ }\textbf {\bibinfo {volume}
  {32}},\ \bibinfo {pages} {5107} (\bibinfo {year} {1985})}\BibitemShut
  {NoStop}%
\bibitem [{\citenamefont {Mizumaki}\ \emph {et~al.}(2005)\citenamefont
  {Mizumaki}, \citenamefont {Uozumi}, \citenamefont {Agui}, \citenamefont
  {Kawamura},\ and\ \citenamefont {Nakazawa}}]{Mizumaki2005}%
  \BibitemOpen
  \bibfield  {author} {\bibinfo {author} {\bibfnamefont {M.}~\bibnamefont
  {Mizumaki}}, \bibinfo {author} {\bibfnamefont {T.}~\bibnamefont {Uozumi}},
  \bibinfo {author} {\bibfnamefont {A.}~\bibnamefont {Agui}}, \bibinfo {author}
  {\bibfnamefont {N.}~\bibnamefont {Kawamura}}, \ and\ \bibinfo {author}
  {\bibfnamefont {M.}~\bibnamefont {Nakazawa}},\ }\bibfield  {title} {\emph
  {\enquote {\bibinfo {title} {Admixture of excited states and ground states of
  a {{Eu}}{\textsuperscript{3+}} ion in \ce{Eu3Fe5O12} by means of magnetic
  circular dichroism},}\ }}\href {\doibase 10.1103/PhysRevB.71.134416}
  {\bibfield  {journal} {\bibinfo  {journal} {Phys. Rev. B}\ }\textbf {\bibinfo
  {volume} {71}},\ \bibinfo {pages} {134416} (\bibinfo {year}
  {2005})}\BibitemShut {NoStop}%
\bibitem [{\citenamefont {Concas}\ \emph {et~al.}(2011)\citenamefont {Concas},
  \citenamefont {Dewhurst}, \citenamefont {Sanna}, \citenamefont {Sharma},\
  and\ \citenamefont {Massidda}}]{Concas2011}%
  \BibitemOpen
  \bibfield  {author} {\bibinfo {author} {\bibfnamefont {G.}~\bibnamefont
  {Concas}}, \bibinfo {author} {\bibfnamefont {J.~K.}\ \bibnamefont
  {Dewhurst}}, \bibinfo {author} {\bibfnamefont {A.}~\bibnamefont {Sanna}},
  \bibinfo {author} {\bibfnamefont {S.}~\bibnamefont {Sharma}}, \ and\ \bibinfo
  {author} {\bibfnamefont {S.}~\bibnamefont {Massidda}},\ }\bibfield  {title}
  {\emph {\enquote {\bibinfo {title} {Anisotropic exchange interaction between
  nonmagnetic europium cations in \ce{Eu2O3}},}\ }}\href {\doibase
  10.1103/PhysRevB.84.014427} {\bibfield  {journal} {\bibinfo  {journal} {Phys.
  Rev. B}\ }\textbf {\bibinfo {volume} {84}},\ \bibinfo {pages} {014427}
  (\bibinfo {year} {2011})}\BibitemShut {NoStop}%
\bibitem [{\citenamefont {Carra}\ \emph {et~al.}(1993)\citenamefont {Carra},
  \citenamefont {Thole}, \citenamefont {Altarelli},\ and\ \citenamefont
  {Wang}}]{Carra1993}%
  \BibitemOpen
  \bibfield  {author} {\bibinfo {author} {\bibfnamefont {P.}~\bibnamefont
  {Carra}}, \bibinfo {author} {\bibfnamefont {B.~T.}\ \bibnamefont {Thole}},
  \bibinfo {author} {\bibfnamefont {M.}~\bibnamefont {Altarelli}}, \ and\
  \bibinfo {author} {\bibfnamefont {X.}~\bibnamefont {Wang}},\ }\bibfield
  {title} {\emph {\enquote {\bibinfo {title} {X-ray circular dichroism and
  local magnetic fields},}\ }}\href {\doibase 10.1103/PhysRevLett.70.694}
  {\bibfield  {journal} {\bibinfo  {journal} {Phys. Rev. Lett.}\ }\textbf
  {\bibinfo {volume} {70}},\ \bibinfo {pages} {694} (\bibinfo {year}
  {1993})}\BibitemShut {NoStop}%
\bibitem [{\citenamefont {Crocombette}\ \emph {et~al.}(1996)\citenamefont
  {Crocombette}, \citenamefont {Thole},\ and\ \citenamefont
  {Jollet}}]{Crocombette1996}%
  \BibitemOpen
  \bibfield  {author} {\bibinfo {author} {\bibfnamefont {J.~P.}\ \bibnamefont
  {Crocombette}}, \bibinfo {author} {\bibfnamefont {B.~T.}\ \bibnamefont
  {Thole}}, \ and\ \bibinfo {author} {\bibfnamefont {F.}~\bibnamefont
  {Jollet}},\ }\bibfield  {title} {\emph {\enquote {\bibinfo {title} {The
  importance of the magnetic dipole term in magneto-circular x-ray absorption
  dichroism for 3d transition metal compounds},}\ }}\href {\doibase
  10.1088/0953-8984/8/22/013} {\bibfield  {journal} {\bibinfo  {journal} {J.
  Phys.: Condens. Matter}\ }\textbf {\bibinfo {volume} {8}},\ \bibinfo {pages}
  {4095} (\bibinfo {year} {1996})}\BibitemShut {NoStop}%
\bibitem [{\citenamefont {Wu}\ and\ \citenamefont {Freeman}(1994)}]{Wu1994}%
  \BibitemOpen
  \bibfield  {author} {\bibinfo {author} {\bibfnamefont {R.}~\bibnamefont
  {Wu}}\ and\ \bibinfo {author} {\bibfnamefont {A.~J.}\ \bibnamefont
  {Freeman}},\ }\bibfield  {title} {\emph {\enquote {\bibinfo {title}
  {Limitation of the {{Magnetic}}-{{Circular}}-{{Dichroism Spin Sum Rule}} for
  {{Transition Metals}} and {{Importance}} of the {{Magnetic Dipole Term}}},}\
  }}\href {\doibase 10.1103/PhysRevLett.73.1994} {\bibfield  {journal}
  {\bibinfo  {journal} {Phys. Rev. Lett.}\ }\textbf {\bibinfo {volume} {73}},\
  \bibinfo {pages} {1994} (\bibinfo {year} {1994})}\BibitemShut {NoStop}%
\bibitem [{\citenamefont {Wachter}(1972)}]{Wachter1972}%
  \BibitemOpen
  \bibfield  {author} {\bibinfo {author} {\bibfnamefont {P.}~\bibnamefont
  {Wachter}},\ }\bibfield  {title} {\emph {\enquote {\bibinfo {title} {The
  optical electrical and magnetic properties of the europium chalcogenides and
  the rare earth pnictides},}\ }}\href {\doibase 10.1080/10408437208244865}
  {\bibfield  {journal} {\bibinfo  {journal} {C R C Crit. Rev. Solid State
  Sci.}\ }\textbf {\bibinfo {volume} {3}},\ \bibinfo {pages} {189} (\bibinfo
  {year} {1972})}\BibitemShut {NoStop}%
\bibitem [{\citenamefont {Lettieri}\ \emph {et~al.}(2003)\citenamefont
  {Lettieri}, \citenamefont {Vaithyanathan}, \citenamefont {Eah}, \citenamefont
  {Stephens}, \citenamefont {Sih}, \citenamefont {Awschalom}, \citenamefont
  {Levy},\ and\ \citenamefont {Schlom}}]{Lettieri2003}%
  \BibitemOpen
  \bibfield  {author} {\bibinfo {author} {\bibfnamefont {J.}~\bibnamefont
  {Lettieri}}, \bibinfo {author} {\bibfnamefont {V.}~\bibnamefont
  {Vaithyanathan}}, \bibinfo {author} {\bibfnamefont {S.~K.}\ \bibnamefont
  {Eah}}, \bibinfo {author} {\bibfnamefont {J.}~\bibnamefont {Stephens}},
  \bibinfo {author} {\bibfnamefont {V.}~\bibnamefont {Sih}}, \bibinfo {author}
  {\bibfnamefont {D.~D.}\ \bibnamefont {Awschalom}}, \bibinfo {author}
  {\bibfnamefont {J.}~\bibnamefont {Levy}}, \ and\ \bibinfo {author}
  {\bibfnamefont {D.~G.}\ \bibnamefont {Schlom}},\ }\bibfield  {title} {\emph
  {\enquote {\bibinfo {title} {Epitaxial growth and magnetic properties of
  {{EuO}} on (001) {{Si}} by molecular-beam epitaxy},}\ }}\href {\doibase
  10.1063/1.1593832} {\bibfield  {journal} {\bibinfo  {journal} {Appl. Phys.
  Lett.}\ }\textbf {\bibinfo {volume} {83}},\ \bibinfo {pages} {975} (\bibinfo
  {year} {2003})}\BibitemShut {NoStop}%
\bibitem [{\citenamefont {W{\"a}ckerlin}\ \emph {et~al.}(2015)\citenamefont
  {W{\"a}ckerlin}, \citenamefont {Donati}, \citenamefont {Singha},
  \citenamefont {Baltic}, \citenamefont {Uldry}, \citenamefont {Delley},
  \citenamefont {Rusponi},\ and\ \citenamefont {Dreiser}}]{Wackerlin2015}%
  \BibitemOpen
  \bibfield  {author} {\bibinfo {author} {\bibfnamefont {C.}~\bibnamefont
  {W{\"a}ckerlin}}, \bibinfo {author} {\bibfnamefont {F.}~\bibnamefont
  {Donati}}, \bibinfo {author} {\bibfnamefont {A.}~\bibnamefont {Singha}},
  \bibinfo {author} {\bibfnamefont {R.}~\bibnamefont {Baltic}}, \bibinfo
  {author} {\bibfnamefont {A.-C.}\ \bibnamefont {Uldry}}, \bibinfo {author}
  {\bibfnamefont {B.}~\bibnamefont {Delley}}, \bibinfo {author} {\bibfnamefont
  {S.}~\bibnamefont {Rusponi}}, \ and\ \bibinfo {author} {\bibfnamefont
  {J.}~\bibnamefont {Dreiser}},\ }\bibfield  {title} {\emph {\enquote {\bibinfo
  {title} {Strong antiferromagnetic exchange between manganese phthalocyanine
  and ferromagnetic europium oxide},}\ }}\href {\doibase 10.1039/C5CC01823D}
  {\bibfield  {journal} {\bibinfo  {journal} {Chem. Commun.}\ }\textbf
  {\bibinfo {volume} {51}},\ \bibinfo {pages} {12958} (\bibinfo {year}
  {2015})}\BibitemShut {NoStop}%
\bibitem [{\citenamefont {Nereson}\ \emph {et~al.}(1964)\citenamefont
  {Nereson}, \citenamefont {Olsen},\ and\ \citenamefont
  {Arnold}}]{Nereson1964}%
  \BibitemOpen
  \bibfield  {author} {\bibinfo {author} {\bibfnamefont {N.~G.}\ \bibnamefont
  {Nereson}}, \bibinfo {author} {\bibfnamefont {C.~E.}\ \bibnamefont {Olsen}},
  \ and\ \bibinfo {author} {\bibfnamefont {G.~P.}\ \bibnamefont {Arnold}},\
  }\bibfield  {title} {\emph {\enquote {\bibinfo {title} {Magnetic
  {{Structure}} of {{Europium}}},}\ }}\href {\doibase 10.1103/PhysRev.135.A176}
  {\bibfield  {journal} {\bibinfo  {journal} {Phys. Rev.}\ }\textbf {\bibinfo
  {volume} {135}},\ \bibinfo {pages} {A176} (\bibinfo {year}
  {1964})}\BibitemShut {NoStop}%
\bibitem [{\citenamefont {Jensen}\ and\ \citenamefont
  {Mackintosh}(1991)}]{Jensen1991}%
  \BibitemOpen
  \bibfield  {author} {\bibinfo {author} {\bibfnamefont {J.}~\bibnamefont
  {Jensen}}\ and\ \bibinfo {author} {\bibfnamefont {A.~R.}\ \bibnamefont
  {Mackintosh}},\ }\href@noop {} {\emph {\bibinfo {title} {Rare Earth
  Magnetism}}}\ (\bibinfo  {publisher} {{Clarendon Oxford}},\ \bibinfo {year}
  {1991})\BibitemShut {NoStop}%
\bibitem [{\citenamefont {Suematsu}\ \emph {et~al.}(1983)\citenamefont
  {Suematsu}, \citenamefont {Ohmatsu}, \citenamefont {Sakakibara},
  \citenamefont {Date},\ and\ \citenamefont {Suzuki}}]{Suematsu1983}%
  \BibitemOpen
  \bibfield  {author} {\bibinfo {author} {\bibfnamefont {H.}~\bibnamefont
  {Suematsu}}, \bibinfo {author} {\bibfnamefont {K.}~\bibnamefont {Ohmatsu}},
  \bibinfo {author} {\bibfnamefont {T.}~\bibnamefont {Sakakibara}}, \bibinfo
  {author} {\bibfnamefont {M.}~\bibnamefont {Date}}, \ and\ \bibinfo {author}
  {\bibfnamefont {M.}~\bibnamefont {Suzuki}},\ }\bibfield  {title} {\emph
  {\enquote {\bibinfo {title} {Magnetic properties of europium-graphite
  intercalation compound {{C}}{\textsubscript{6}}{{Eu}}},}\ }}\href {\doibase
  10.1016/0379-6779(83)90005-X} {\bibfield  {journal} {\bibinfo  {journal}
  {Synth. Met.}\ }\textbf {\bibinfo {volume} {8}},\ \bibinfo {pages} {23}
  (\bibinfo {year} {1983})}\BibitemShut {NoStop}%
\bibitem [{\citenamefont {Averyanov}\ \emph {et~al.}(2016)\citenamefont
  {Averyanov}, \citenamefont {Tokmachev}, \citenamefont {Karateeva},
  \citenamefont {Karateev}, \citenamefont {Lobanovich}, \citenamefont
  {Prutskov}, \citenamefont {Parfenov}, \citenamefont {Taldenkov},
  \citenamefont {Vasiliev},\ and\ \citenamefont {Storchak}}]{Averyanov2016}%
  \BibitemOpen
  \bibfield  {author} {\bibinfo {author} {\bibfnamefont {D.~V.}\ \bibnamefont
  {Averyanov}}, \bibinfo {author} {\bibfnamefont {A.~M.}\ \bibnamefont
  {Tokmachev}}, \bibinfo {author} {\bibfnamefont {C.~G.}\ \bibnamefont
  {Karateeva}}, \bibinfo {author} {\bibfnamefont {I.~A.}\ \bibnamefont
  {Karateev}}, \bibinfo {author} {\bibfnamefont {E.~F.}\ \bibnamefont
  {Lobanovich}}, \bibinfo {author} {\bibfnamefont {G.~V.}\ \bibnamefont
  {Prutskov}}, \bibinfo {author} {\bibfnamefont {O.~E.}\ \bibnamefont
  {Parfenov}}, \bibinfo {author} {\bibfnamefont {A.~N.}\ \bibnamefont
  {Taldenkov}}, \bibinfo {author} {\bibfnamefont {A.~L.}\ \bibnamefont
  {Vasiliev}}, \ and\ \bibinfo {author} {\bibfnamefont {V.~G.}\ \bibnamefont
  {Storchak}},\ }\bibfield  {title} {\emph {\enquote {\bibinfo {title}
  {Europium {{Silicide}} \textendash{} a {{Prospective Material}} for
  {{Contacts}} with {{Silicon}}},}\ }}\href {\doibase 10.1038/srep25980}
  {\bibfield  {journal} {\bibinfo  {journal} {Sci Rep}\ }\textbf {\bibinfo
  {volume} {6}},\ \bibinfo {pages} {25980} (\bibinfo {year}
  {2016})}\BibitemShut {NoStop}%
\end{thebibliography}%

\end{document}